\documentclass[aps,prl,twocolumn,superscriptaddress]{revtex4}
\usepackage{amsfonts,amssymb,amsmath}
\usepackage{color}
\usepackage{graphicx}
\usepackage{dcolumn}
\usepackage{bm}

\begin{document}

\title{Unconventional continuous phase transition in a three
  dimensional dimer model}

\author{Fabien Alet}
\email{alet@irsamc.ups-tlse.fr}
\affiliation{Laboratoire de Physique Th\'eorique, UMR CNRS 5152, Universit\'e Paul Sabatier, 31062 Toulouse, France}
\affiliation{Service de Physique Th\'eorique, URA CNRS 2306, CEA Saclay, 91191 Gif sur Yvette, France}

\author{Gr\'egoire Misguich}
\author{Vincent Pasquier}
\affiliation{Service de Physique Th\'eorique, URA CNRS 2306, CEA
  Saclay, 91191 Gif sur Yvette, France}
\author{Roderich Moessner}
\affiliation{Laboratoire de Physique Th\'eorique, UMR CNRS 8549, \'Ecole Normale
  Sup\'erieure, 75005 Paris, France}
\author{Jesper Lykke Jacobsen}
\affiliation{LPTMS, UMR CNRS 8626, Universit\'e Paris Sud, 91405 Orsay, France}
\affiliation{Service de Physique Th\'eorique, URA CNRS 2306, CEA Saclay, 91191 Gif sur Yvette, France}

\date{\today}

\begin{abstract}
Phase transitions occupy a central role in physics, due both to their
experimental ubiquity and their fundamental conceptual importance.
The explanation of universality at phase transitions was the great success
of the theory formulated by Ginzburg and Landau, and extended through the
renormalization group by Wilson. However, recent theoretical suggestions
have challenged  this point of  view  in  certain situations. In this Letter we
report the first large-scale simulations  of a three-dimensional model
proposed  to be a  candidate  for requiring  a description  beyond the
Landau-Ginzburg-Wilson framework: we study  the phase  transition from
the dimer crystal to  the Coulomb phase in the  cubic dimer model. Our
numerical results strongly  indicate that the transition is continuous
and are  compatible with a  {\it  tricritical} universality  class, at
variance with previous proposals.

\end{abstract}


\maketitle

The   Landau-Ginzburg-Wilson         (LGW)       theory  of      phase
transitions~\cite{textbook} has been a remarkably powerful approach to
study critical phenomena, both  in classical and quantum  systems. The
basic assumption is that each phase is characterized by its symmetries
and a local  order parameter for the broken  symmetries (in case of an
ordered phase). To describe a  phase transition, this approach amounts
to  expanding the free  energy   of the  system  in powers  of  the order
parameter(s) describing  the ordered phase(s). Exceptions  are however
known in 2d classical or equivalently 1d  quantum systems where stable
{\em critical phases} (with power-law correlations but {\em no} broken
symmetry) exist, the  low temperature phase  of the 2d XY  model being
one  of the  most famous examples.  The associated Kosterlitz-Thouless
(KT)  phase transition~\cite{KT} and   the role of topological defects
have become very  fruitful concepts in  the statistical physics  of 2d
phases not described by a simple order parameter in the LGW sense.

The stimulating  idea that ``non-LGW''  continuous phase transitions could
also occur in higher dimensions has recently been proposed in the context
of quantum magnetism~\cite{senthil}.  There, the possibility
of a  generic ({\it i.e.} not requiring fine tuning) continuous transition
between two   phases  with different symmetry   breakings  (N\'eel and
Valence Bond Solid states) was pointed out,  in contrast to the
LGW prediction of either an intermediate phase with none of the two orders, 
or a coexistence region, or a direct first order transition. So far, 
simulations on all candidate quantum models~\cite{numeric} rather see a
first order phase transition, a possibility which can never be discarded for a given
microscopic model (see, however, recent claim of a continuous transition~\cite{numeric2}).

From  these perspectives, {\it classical}  dimers  at close packing on
simple hypercubic lattices are particularly interesting as they are too
constrained to form a liquid with a finite correlation length, even at
infinite  temperature: instead  of decaying exponentially, the dimer-dimer
correlations are algebraic,
both in 2d~\cite{alet05} and  3d~\cite{huse}. In 2d, the   transition
from such a critical state to a  broken symmetry phase (dimer crystal)
has been  studied  in  Ref.~\onlinecite{alet05}. In this Letter we study the
analogous  transition in   a three-dimensional  classical dimer model.
Both   in  2d and    3d, a  naive  LGW  expansion   in  terms  of  the
low-temperature order parameter  completely misses the critical nature
of  the high-temperature phase and  thus cannot describe correctly the
transition.  Whereas in  2d the situation  is well understood in terms
of a KT transition~\cite{alet05},  our numerical simulations on the 3d
model show the existence  of  a single  continuous {\it
second-order} phase  transition between the  algebraic liquid  at high
temperature  (so-called Coulomb  phase)   and a  crystal  with  broken
lattice symmetries. Our high-precision  Monte Carlo (MC) data allow 
independently to locate the transition    starting from both   phases. We
determine  the  critical exponents and find that  they are
surprisingly close to those of a tricritical  point. In a related
recent work, Bergman {\it et al.}~\cite{Bergman} further argue that the
very existence  of  the Coulomb  phase guarantees  that this transition lies outside the standard LGW  framework (see also
Ref.~\onlinecite{ms04}). The Coulomb phase
in   turn   owes its  presence   to   the  absence of   unpaired sites
(monomers)~\cite{huse}. In this spirit, it  is the suppression of such
topological   defects  which opens the  way  to  new types of critical
behavior. Indeed, such   ``topological engineering''  has  previously
been used  in the context of  liquid crystals~\cite{Lammert}, and more
recently for a non-linear $\sigma$ model~\cite{mw};  as we discuss before
concluding, the numerical values of the critical exponents obtained in
Ref.~\onlinecite{mw} are not consistent with the ones reported here.

The model     is a    3d   extension  of      the   one  studied    in
Ref.~\onlinecite{alet05}  on the square  lattice.   Configurations are
dimer coverings of  the simple cubic lattice  of volume $N=L^3$ sites,
with $L$ the linear dimension.  Dimers are hard-core and close-packed,
{\em i.e.}
every lattice site is part of one and only one dimer. Interactions
favor aligment of nearest neighbours (n.n.) dimers on plaquettes of the lattice:
\begin{equation} 
\label{eq:model}
E = - \sum_{\rm plaquettes} n_{||} + n_{=} + n_{//},
\end{equation}
with $n_{||}, n_{=}$ and $n_{//}$ denoting the number of plaquettes with
parallel n.n. dimers in the $x,y$ and $z$ directions.
Simulations (up  to $N=96^3$) are performed with a recent MC directed-loop
algorithm~\cite{sandvik}. 

At  $T=0$, the  dimers  order  in  columns  to  minimize   the energy,
resulting in  a 6-fold degenerate ground  state.  The associated order
parameter      is        a     3-component     vector   $m^\alpha({\bf
r})=(-)^{r_\alpha}n_\alpha({\bf r})$, with $n_\alpha({\bf r})=1$ for a
dimer  pointing in direction $\alpha \in  {x,y,z}$ at  site ${\bf r}$,
and $0$  otherwise.  Naively, one would  expect a  high-$T$ phase with
$\langle    \vec  m\rangle=0$   and   exponentially   decaying   dimer
correlations.  However, as shown by  Huse {\it et al.}~\cite{huse}, at
$T=\infty$ the system is in a ``Coulomb  phase'', with no true long-range
order  but  with dipolar dimer-dimer  correlations.   To see this, the
appropriate  variable   is  the  ``electric        field''~\cite{huse}
$E_\alpha({\bf r})=(-)^{\bf    r}  n_\alpha({\bf   r})$.   This  field
satisfies ${\bf \nabla} \cdot {\bf E} =  (-)^{\bf r} =  \pm 1$, as the
dimers are close-packed. The Coulomb phase can be characterized~\cite{huse}
in   the   continuum    by  an effective     ``electrostatic''  action
$S=\frac{K}{2}\int d{\bf r}  {\bf  E}^2({\bf r})$ which generates  the
dipolar  correlations. Dimer   fluxes   $\phi={\bf  \int}_\Sigma  {\bf
E}\cdot d{\bf    S}$ through the  planes   perpendicular  to the units
vectors  are  conserved quantities and  vanish  on average. One easily
shows that flux fluctuations allow to calculate $K$ \begin{equation}
\label{eq:K}
\left< \phi^2 \right> / L = \frac{1}{3 L}(\left< \phi_x^2 \right> + \left<
\phi_y^2 \right> + \left< \phi_z^2 \right>) = 1/K.
\end{equation}

A close similarity with a  3d XY model  can be seen through a  duality
transformation~\cite{kogut}   in which ${\bf \nabla}   \cdot {\bf E} =
(-)^{\bf r}$ is enforced by an angular Lagrange multiplier $\theta$ at
each site. The  discrete sums on $E_\alpha$  are  then performed  by a
Poisson formula, resulting in  an XY interaction (Villain form) between
the  $\theta$  variables.     In this language,    the   Coulomb phase
corresponds to an  ordered  phase  with  broken $O(2)$ symmetry    for
$\theta$, and $1/K$ is the associated spin stiffness.

\begin{figure}
\includegraphics[width=8cm]{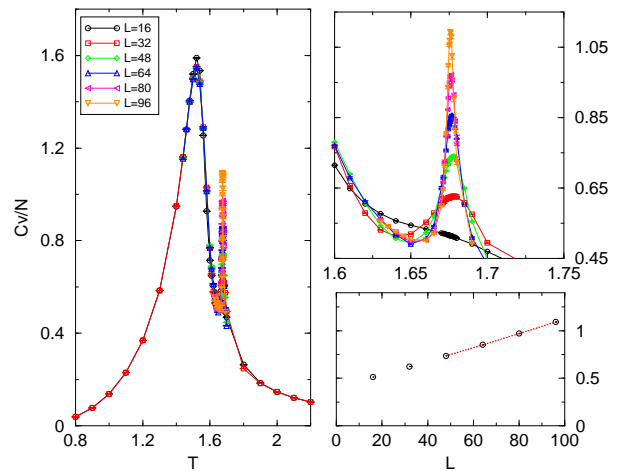}
\caption{Left panel: Specific heat per site $C_v/N$ as function of temperature
  $T$ for different system sizes $L$. Top right panel: Zoom on the second
  peak. Bottom right panel: Scaling of the specific heat at
  $T_c^{C_v}=1.676$ as a function of system size $L$. The dotted line
  denotes a power-law fit for the $4$ largest systems  $L=48,64,80,96$ (see text).}
\label{fig:Cv}
\end{figure}

We first present thermodynamic results. Fig.~\ref{fig:Cv} (left panel)
shows the behavior of the specific heat per site $C_v/N$ as a function
of $T$. Two close-by peaks are observed around $T\sim 1.52$ and $T\sim
1.67$. The first peak is much broader and {\it  does not} diverge with
system size: since it is already present and almost converged on small
lattices $L<16$ (not shown), it  cannot be associated to
any long distance or critical behavior.  The second  peak is much more
characteristic of a phase  transition:  it diverges  with $L$,  with a
power-law-like envelope typical  of second order phase transitions (see
top right  panel). Note  that this  peak  is extremely hard  to detect
since it is absent on small systems ($L\leq 16$) and also very sharp. 
We interpret this peak as the signature of the direct transition (see
below) between the Coulomb and columnar phase. Our best estimate for the 
temperature of its divergence is $T_c^{C_v}=1.676(1)$.  To determine the  
nature of the transition, we also considered energy histograms and the energy
cumulant~\cite{challa}  defined  as $1-\left<  E^4 \right>/3\left< E^2
\right>^2$. No sign of double peak is detected in histograms and the
energy cumulant saturates to $2/3$ at the transition point:
this indicates that  {\it the transition  is not first order}. 

\begin{figure}
\includegraphics[width=8cm]{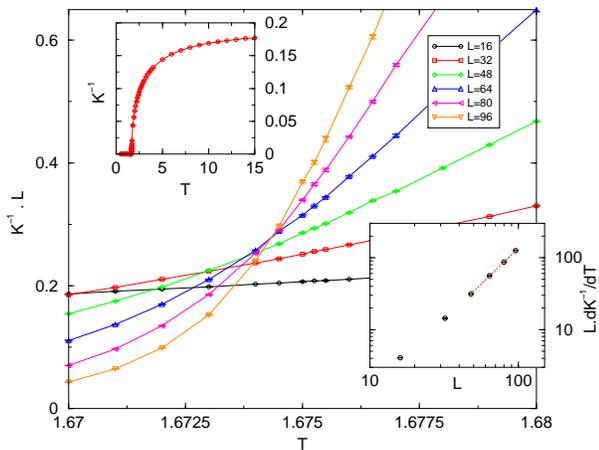}
\caption{
   Stiffness $K^{-1}$ multiplied by  $L$ (obtained from $\left< \phi^2
   \right>$, see  Eq.~\ref{eq:K}) versus $T$  near the transition, for
   different system sizes $L$.  Left inset: Stiffness $K^{-1}$ for the
   whole high-$T$  range   for $L=32$.   Right inset:   Scaling of the
   derivative  $LdK^{-1}/dT$ versus  $L$   in log-log  scale  for  the
   estimated  critical temperature  $T_c^K=1.6745$.   The  dotted line
   denotes a power-law fit for the $4$ largest systems.}
\label{fig:Rho}
\end{figure}

\begin{figure}
\includegraphics[width=8cm]{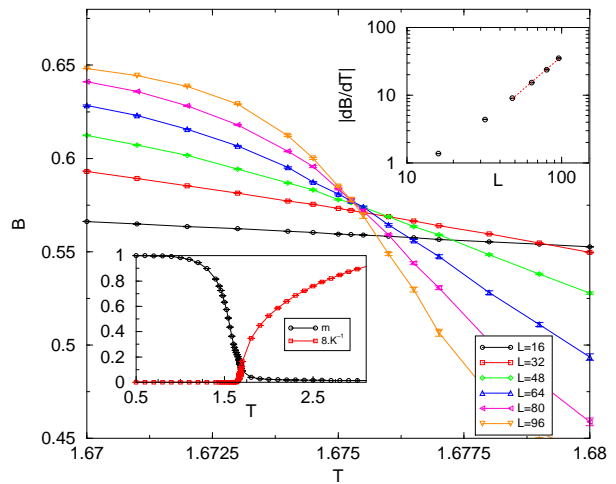}
\caption{Crossing of the columnar Binder cumulant $B$ as a function of $T$ near
the transition point for different system sizes $L$. The sample $L=16$ is
again out of scaling. Left inset: Columnar order parameter $m$ and stiffness $K^{-1}$ (multiplied by 8) in the
whole $T$ range for $L=32$. Right inset: Cumulant derivative
$|dB/dT|$ at $T_c^{\rm col}=1.67525$ versus $L$ in log-log scale. The dotted
line is a power-law fit for the $4$ largest L.}
\label{fig:SachdevCum}
\end{figure}

Let us now consider the high-$T$ phase. The left inset of Fig.~\ref{fig:Rho} displays typical data for $K^{-1}$ for a
sample $L=32$. $K$ is finite in the whole high-T range
(with a value $K(T=\infty)=5.12(1)$ in agreement with Ref.~\onlinecite{huse})
and diverges below $T\simeq  1.6$ (rough   estimate    from  the  plot,  and
inset   of Fig.~\ref{fig:SachdevCum}). We expect for a second order phase
transition a scaling form of the type
\begin{equation}
K^{-1} = \left< \phi^2 \right> / L = L^{-z} f ( L^{1/\nu}.(T-T_c))
\end{equation}
where $z$ is a scale exponent,  $\nu$ the correlation length exponent,
$f$ a scaling function and $T_c$ the critical temperature. Dimensional
analysis of the Coulombian action gives $z=1$ and therefore, at $T_c$,
the  curves  of $LK^{-1}$ intersect   for all $L$  and  the derivative
$L\frac{dK^{-1}}{dT}$ scales  as $L^{1/\nu}$.  Numerically an accurate
crossing  point (see  main   panel  of Fig.~\ref{fig:Rho})  is  indeed
obtained for $z=1$,  which is also   a good check of the  second-order
nature of the   transition.  An  estimate  $T_c^{K}=1.6745(5)$ can  be
obtained from the crossing of the curves for  the largest $L$.  In the
Coulomb   phase,   dimer-dimer  correlations   are  expected   to   be
dipolar~\cite{huse}, and this is found to be indeed true all along the
high-$T$ phase.  The prefactor in the dipolar  form of the correlation
functions  varies  as $1/K$, and  we  have checked with high precision
that  the value  of  $K$  obtained  from flux  fluctuations  perfectly
coincides  with   that  from   correlations.   {\it   Monomer-monomer}
correlators         are       also         available       in      the
simulations~\cite{sandvik,alet05}  and we find  that test monomers are
deconfined   from  $T=\infty$  down    to  $T_c$,  confirming the Coulombian  nature of
the phase.

To  probe the nature  of the low-$T$  phase, we calculate the columnar
order parameter
\begin{equation}
m=\frac{2}{N}||\sum_{\bf r}\vec m({\bf r})  ||
\end{equation} and its Binder
cumulant~\cite{Binder} $B=1-\left< m^4 \right>  /3 \left<  m^2 \right>
^2$. The left  inset of Fig.~\ref{fig:SachdevCum} shows  the expectation
value $\left<m\right>$ for a sample $L=32$ (for illustration, $K^{-1}$
is again shown). Columnar  order is observed to set in
at low $T$. Binder cumulants in the main panel  admit a crossing point
for systems   with  different $L$,  leading   to an estimate $T_c^{\rm
col}=1.67525(50)$. Assuming   the   standard  scaling form   $B    = f  (
L^{1/\nu}.(T-T_c))$, the derivative $dB/dT$ should scale as $L^{1/\nu}$ at
$T_c$.

The previous findings and the  agreement between the various estimates
of $T_c$   clearly indicate  that the  model  displays  a {\it  single
second-order phase transition  between Coulomb and columnar phases}. A
straightforward choice for a LGW theory would be to use $m$ as an order
parameter.  However, this fails as  the Coulomb phase is  not just a
simple liquid where all correlations  decay exponentially. Rather,  it
retains algebraic dimer  correlations whose dipolar nature, crucially,
does not  lead to a  peak in the  structure factor anywhere in Fourier
space   (unlike the analogous  situation in 2d~\cite{sandvik,alet05}).
Indeed,  the natural    variable  with algebraic   correlations  is  a
coarse-grained ``electric'' field $E$; however, this variable exhibits
no long-range  order   in {\it  either}  phase.  Instead,   it is  the
fluctuations of $E$  that  distinguish the  two  phases.  It is   also
instructive to examine the transition  from the Coulomb side with  the
dual angles $\theta$. It allows to map the dimer  problem onto a model
of interacting vortex loops  with a long-range $1/r$ potential.  While
these loops   are  dilute   in  the   Coulomb  phase,  they have    to
``proliferate''  to reproduce  a  low-temperature crystal  phase  with
frozen dimer   positions  (thus   highly fluctuating   dual  variables
$\theta$).  Intuitively, the restoration of  the $O(2)$ symmetry would
be   through an   inverted   3d  $XY$   transition,  which  is  however
incompatible   with  the  critical exponents   found  numerically (see
below). The  crucial difference with a simple  O(2) spin model  can be
traced back to the background electric charges $\pm 1$ which couple to
the vortex loops and presumably  affect  their proliferation. We  note
the  similar  analysis   of   Ref.~\onlinecite{Bergman},  where     an
unconventional  non-LGW transition is  predicted  in a closely related
model.

We now    come to  the universality  class   of the   transition.  The
correlation  length exponent $\nu$ can   be extracted from the scaling
with  $L$ of   stiffness  $LdK^{-1}/dT$  or  Binder  cumulant  $dB/dT$
derivatives  at the critical  temperature $T_c$, which can be calculated 
thermodynamically in the MC process. Taking  into account only the
largest     $L\geq   48$  (see    insets   of  Figs.~\ref{fig:Rho} and
\ref{fig:SachdevCum}),    we   obtain compatible   estimates $\nu^{\rm
K}=0.50(4)$ and $\nu^{\rm col}=0.51(3)$  (error bars take into account
stability of fits toward inclusion of smaller samples and uncertainty
on $T_c$).   The specific  heat   critical  exponent $\alpha$   can be
extracted from its scaling at the critical  point: $C_v(T_c)/N = c_0 +
A L^{\alpha/\nu}$ where $A$ is  a constant and  $c_0$ the regular part
at the transition  ($c_0$  is non negligible  as  can be seen  for the
$L=16$   sample in Fig.~\ref{fig:Cv}). A fit for the  largest $L$ (see
inset in  Fig.~\ref{fig:Cv})  gives  $\alpha/\nu=1.11(15)$, leading  to
$\alpha=0.56(7)$.   Hyperscaling $\alpha=2-\nu  d$ is  thus  satisfied
within error bars. The last  independent exponent can be obtained from
the scaling of the columnar order parameter at criticality $m(T_c)\sim
L^{-\beta/\nu}$ or from the associated susceptibility $\chi = \left< m^2
\right>  -\left< m  \right>   ^2\sim L^{\gamma/\nu}$.  Using  standard
relations between critical  exponents,  we obtain for the  correlation
function exponent  $\eta=-0.02(5)$.
This   set  of  exponents  excludes  some    simple  $3d$
universality classes  (such as  $O(2)$,  $O(3)$ with or without  cubic
anisotropy)  but  are compatible  with  the universality   class of an
$O(n)$ {\it tricritical point} at  its upper critical  dimension $d=3$ for which $\nu=\alpha=1/2$ and
$\eta=0$ (up to logarithmic corrections).  We also note that the value
at $T_c$ of the  cumulant of the 3-dimensional  order parameter $m$ is
compatible within error bars (see   Fig.~\ref{fig:SachdevCum}) with  the value $0.56982
\ldots$ of  a tricritical  $O(3)$ theory (for $d\geq3$)~\cite{brezin}.
With dimers, we do not have direct access  to the XY order parameter
$\vec   n(r)$   for the dual    angles   $\theta$.  We  can  however
investigate  the fluctuations of  the  electric flux.  $\phi_z$ is the
integer-valued Noether   charge  associated  to  the $O(2)$   symmetry
(``total   angular momentum'' if  $z$  is  interpreted as 
the  time direction):
$\phi_z=\kappa\int dx  dy (n_1\partial_z n_2-n_2\partial_z n_1)$ where
$\kappa$        appears          in     the       ``kinetic''     term
$\frac{\kappa}{2}(\partial_z\vec{n})^2$  of    the dual $O(2)$ action.
Above  $T_c$ the  typical  flux scales  as  $\sqrt  L$  and the ratios
$\langle\phi_z^4\rangle/\langle\phi_z^2\rangle^2$                  and
$\langle\vec\phi^4\rangle/\langle\vec\phi^2\rangle^2$ are respectively
equal   to    3 and   5/3,  in   agreement   with  the   Gaussian  and
$O(3)$-symmetric nature of fluctuations  in  the Coulomb phase.  We
believe  that the distribution of $\phi$  is universal at $T_c$ and we
measured $\langle\phi^2\rangle=0.28(2)$               and
$\langle\phi^4\rangle=0.25(4)$.   
The smallness of these quantities at the critical point means that
the discrete nature  of  $\vec\phi=(\phi_x,\phi_y,\phi_z)$  cannot be
neglected there.

While it  may  be seen  as the only   way to  reconcile  the numerical
results  with a LGW analysis,  the tricritical  universality class is
rather unexpected  here as it would  imply a ``hidden'' fine-tuning of
parameters of  the effective action.  It  is  also quite possible that
the  exponents found   are close,   but not   equal,  to  tricritical
exponents,    thereby          defining               a            new
``non-LGW''~\cite{Bergman,senthil} universality  class.  The
absence  of  monomers is  in   fact very similar   to  the  absence of
``hedgehogs''    in   the    models   studied   by     Motrunich   and
Vishwanath~\cite{mw}, which also display a  transition
from a broken-symmetry phase   to  a Coulomb liquid.   However,   our
critical exponents do not   match those of Ref.~\onlinecite{mw}.  This
discrepancy  might  be  due  to one   or  several factors:   i)  lattice cubic
anisotropies (not  present  in Ref.~\onlinecite{mw}   but  potentially
relevant)  exist in our model  which admits 6 ground-states related by
cubic symmetry.  ii)  the simulations of  Ref.~\onlinecite{mw} may not
be in the scaling regime.  iii) the  proximity of a tricritical point
(for instance at finite monomer doping),  could affect finite-size
estimations of the exponents and hide the true critical behavior. Finally,
even though we simulate systems with $N$ up to $96^3$, a first-order
transition with a large correlation length can never be excluded from finite-size simulations.

In conclusion,  our study has  established  the crystallization in the
cubic dimer model as an  example of an unconventional phase transition
in a classical  model in 3 dimensions.   The conventional LGW approach
to  phase transitions   is  currently under attack from    many
sides~\cite{senthil,Bergman,mw,bosons},  but shows considerable  resilience   in microscopic models~\cite{numeric}. From  this
perspective, our results are promising since they cannot be described by a strict application of the LGW
scheme~\cite{textbook} of an expansion in terms of the low-T order
parameter. Further analytical calculations and
numerical tests are needed to extend this study, notably  to
investigate the  possible proximity   to a  tricritical  universality
class. This will hopefully allow a deeper  understanding of the limits
of a central concept in statistical physics.

We thank C. Henley, W, Krauth, S. Sondhi for fruitful discussions and D. Huse for
comments on the manuscript. F.A. is supported by the French ANR program. Calculations were performed  on the clusters
Tantale at CCRT/CEA (project    575)  and     Gallega    at   SPhT/CEA
using  the  ALPS libraries~\cite{ALPS}.


\vspace{-0.5cm}

\begin{thebibliography}{99}

\bibitem{textbook} See {\it e.g.} L.D. Landau and E.M. Lifshitz, {\it
  Statistical physics} (Pergamon, New York, 1980) 3rd ed.

\bibitem{KT} J.M. Kosterlitz and D.J. Thouless, J. Phys. C {\bf 6},
  1181 (1973)

\bibitem{senthil} T. Senthil {\it et al.}, Science {\bf 303}, 1490 (2004); Phys. Rev. B {\bf 70}, 144407
  (2004)

\bibitem{numeric} A.W. Sandvik {\it et al.}, Phys. Rev. Lett. {\bf 89},
  247201 (2002); A. Kuklov, N. Prokof'ev and B. Svistunov, {\it ibid} {\bf
  93}, 230402 (2004); L. Spanu, F. Becca and S. Sorella, Phys. Rev. B {\bf
  73}, 134429 (2006); J. Sirker {\it et al.}, {\it ibid} {\bf 73}, 184420
  (2006); A. B. Kuklov {\it et al.}, Ann. Phys. {\bf 321}, 1602 (2006)

\bibitem{numeric2} S. Isakov {\it et al.}, preprint cond-mat/0602430

\bibitem{alet05} F. Alet {\it et al.}, Phys. Rev. Lett. {\bf 94}, 235702 (2005)

\bibitem{Bergman} D.L. Bergman, G.A. Fiete and L. Balents, Phys. Rev. B {\bf 73}, 134402 (2006)

\bibitem{ms04} O.I. Motrunich, and T. Senthil, Phys.  Rev. B {\bf 71}, 125102
(2005)

\bibitem{huse} D. Huse {\it et al.}, Phys. Rev. Lett. {\bf 91}, 167004 (2003)

\bibitem{Lammert} P.E. Lammert, D.S. Rokhsar and J. Toner,
  Phys. Rev. Lett. {\bf 70}, 1650 (1993)

\bibitem{mw} O.I. Motrunich and A. Vishwanath, Phys. Rev. B {\bf 70}, 075104
  (2004)

\bibitem{sandvik} A.W. Sandvik and R. Moessner, Phys. Rev. B {\bf 73}, 144504 (2006) 

\bibitem{kogut} T. Banks, R. Myerson, and J. Kogut, Nucl. Phys. B {\bf 129}, 493 (1977)

\bibitem{challa} M.S.S. Challa, D.P. Landau and K. Binder, Phys. Rev. B {\bf 34}, 1841 (1986)

\bibitem{Binder} K. Binder, Z. Phys. B {\bf 43}, 119 (1981)

\bibitem{brezin} E. Br\'ezin and J. Zinn-Justin, Nucl. Phys. B {\bf 257} 867 (1985)

\bibitem{bosons} L. Balents {\it et al.}, Phys. Rev. B {\bf 71}, 144508
  (2005);  T. Senthil and M.P.A. Fisher, Preprint cond-mat/0510459

\bibitem{ALPS} F. Alet {\it et al.}, J. Phys. Soc. Jap. Suppl. {\bf 74}, 30
  (2005); M. Troyer, B. Ammon and E. Heeb, Lect. Notes Comput. Sci. {\bf
  1505}, 191 (1998);  See {\tt http://alps.comp-phys.org}.
\end{thebibliography}
\end{document}